\documentclass[apj]{emulateapj}

\newcommand {\kms}    {km~s$^{-1}$}

\newcommand {\wlam}   {$W_{\lambda}$} 
\newcommand {\lya}    {Ly$\alpha$}   

\newcommand {\HI}     {\ion{H}{1}}   
\newcommand {\OVI}    {\ion{O}{6}}   
\newcommand {\CIII}   {\ion{C}{3}}   

\newcommand {\NHI}    {$N_{\rm HI}$}
\newcommand {\NOVI}   {$N_{\rm OVI}$}

\newcommand {\NCIV}   {$N_{\rm CIV}$}
\newcommand {\NSiIII} {$N_{\rm SiIII}$}

\newcommand {\NV}     {\ion{N}{5}}
\newcommand {\CIV}    {\ion{C}{4}}
\newcommand {\SiIV}   {\ion{Si}{4}}
\newcommand {\SiIII}  {\ion{Si}{3}}

\newcommand {\FeIII}  {\ion{Fe}{3}}

\newcommand {\etal}  {et~al.}
\newcommand {\cd}    {cm$^{-2}$}

\begin{document}

\title{THE METALLICITY OF INTERGALACTIC GAS IN COSMIC VOIDS}

\author{John T. Stocke, Charles W. Danforth,  J. Michael Shull, Steven V. Penton}
\affil{Center for Astrophysics and Space Astronomy, Department of Astrophysical
and Planetary Sciences, UCB-389, University of Colorado, Boulder, CO 80309}
\author{and Mark L. Giroux}
\affil{Department of Physics, Astronomy, \& Geology, East Tennessee State University, \\ 
  Box 70652, Johnson City, TN 37614}

\email{stocke@casa.colorado.edu, danforth@casa.colorado.edu, 
 mshull@casa.colorado.edu, spenton@casa.colorado.edu, girouxm@etsu.edu}

\shorttitle{Metallicity in Voids}
\shortauthors{Stocke et~al.}

\begin{abstract}

We have used the Hubble/STIS and FUSE archives of ultraviolet spectra of 
bright AGN to identify intergalactic \lya\ absorbers in nearby ($z\leq$0.1) voids. 
From a parent sample of 651 \lya\ absorbers, we identified 61 ``void absorbers'' 
located $>1.4h^{-1}_{70}$~Mpc from the nearest L* or brighter galaxy.
Searching for metal absorption in high-quality (S/N$>$ 10) spectra at the 
location of three diagnostic metal lines (O~VI $\lambda1032$, C~IV $\lambda1548$, 
Si~III $\lambda1206$), we detected no metal lines in any individual 
absorber, or in any group of absorbers using pixel co-addition techniques. 
The best limits on metal-line absorption in voids were set using four strong 
\lya\ absorbers with \NHI\ $>10^{14}$ \cd, with $3\sigma$ equivalent-width limits 
ranging from 8 m\AA\ (\OVI), 7--15 m\AA\ (\CIV), and 4--10 m\AA\ (\SiIII).   
Photoionization modeling yields metallicity limits $Z < 10^{-1.8\pm0.4}~Z_{\odot}$ 
from non-detections of \CIV\ and \OVI, some $\sim 6$ times lower than those
seen in \lya/\OVI\ absorbers at $z < 0.1$.  
Although the void \lya\ absorbers could be pristine material, considerably 
deeper spectra are required to rule out a ``universal metallicity floor"
produced by bursts of early star formation, with no subsequent star formation 
in the voids.   The most consistent conclusion derived from these low-$z$ 
results, and similar searches at $z = 3-5$, is that galaxy filaments have 
increased their mean IGM metallicity by factors of 30--100 since $z \sim 3$. \\

\end{abstract}


\section{INTRODUCTION} 

Enormous spatial voids in the large-scale distribution of galaxies 
were first discovered by Gregory \& Tifft (1976) during 
a redshift survey of the Coma Cluster of galaxies.  Kirshner
et~al. (1981) found an enormous region of extreme galaxy
underdensity in the constellation Bo\"otes, suggesting that voids are
a common feature of intergalactic space.  The existence and
characterization of voids was put on a firm basis by Geller \& Huchra 
(1989) using the Center for Astrophysics Redshift Survey
(CfARS).  Since then, every major galaxy survey (e.g., Sloan Digital
Sky Survey [SDSS], Las Campanas Redshift Survey [LCRS], \ion{H}{1}
surveys of Haynes \& Giovanelli [1985]) has discovered new voids,
extending our understanding of their size and ubiquity.
The SDSS database has also enabled detailed studies of ``void
galaxies'' in the regions of lowest galaxy density.  These regions 
are typically filaments or strings of galaxies that separate immense 
voids into smaller void volumes.  Void galaxies appear to differ in 
luminosity function (fainter $L^*$, Hoyle \etal\ 2005) and other 
properties (e.g., higher star formation rate) from similar type galaxies 
in denser regions (Rojas \etal\ 2004).  On the theoretical side, large
volume numerical simulations have been used to study voids
(Gottl\"ober \etal\ 2003) in a cosmological context, such as estimating 
their spatial densities and the abundance of bound dark-matter halos 
within them.

In this paper, we describe a new approach to studying voids through
the detection of \lya\ absorbing clouds within their volumes.  This
line of research was made possible after the launch of the {\it Hubble 
Space Telescope} (HST) and its UV spectrographs: the {\it Faint Object
Spectrograph} (FOS), the {\it Goddard High Resolution Spectrograph} (GHRS), 
and the {\it Space Telescope Imaging Spectrograph} (STIS).  Far-ultraviolet
spectra of higher Lyman lines and key metal lines (\OVI, \CIII) were 
obtained by FUSE, the {\it Far Ultraviolet Spectroscopic Explorer} 
(Moos \etal\ 2000).  With HST, Morris \etal\ (1991) and Bahcall \etal\ 
(1991) discovered low-$z$ \lya\ 
absorption lines in the spectrum of 3C\,273.  The subsequent HST/FOS Key
Project on QSO Absorption Lines (Weymann \etal\ 1998) found a surprisingly
high line frequency, $d{\cal N}/dz \approx 35$, of strong \lya\ absorbers 
at low redshifts.  The Key Project survey was sensitive to  \lya\ absorbers 
with equivalent widths \wlam\ $\geq 240$ m\AA, corresponding to column 
densities \NHI\ $\geq 10^{14}$ cm$^{-2}$ for $b = 25$ \kms. Higher resolution 
\lya\ studies with GHRS and the STIS/E140M echelle have equivalent-width 
sensitivities (\wlam\ $\geq 15$ m\AA\ or log\,\NHI $\geq 12.44$)
at $z=0$, comparable to recent {\it Keck Observatory} spectra at $z=2-3$.  
At these sensitivities, Penton, Stocke, \& Shull (2004) observed 
$d{\cal N}/dz \approx 200$ \lya\ absorbers per unit redshift. 

Our further efforts on low-$z$ IGM absorbers have examined their metal-line
content (Danforth \& Shull 2005, 2007) as well as the spatial distribution
of galaxies around the \lya\ absorbers. Our early studies (Penton, Shull,
\& Stocke 2000; Penton, Shull, \& Stocke 2000) demonstrated that, on
average, the strong \lya\ absorbers trace the galaxy filaments, while 
most weaker absorbers tend
to be found in voids (Penton, Stocke, \& Shull 2002; McLin, \etal\ 2002). 
Stocke \etal\  (1995) targeted bright QSOs and BL\,Lac Objects behind very 
nearby voids and discovered a few \lya\ absorbers within their boundaries, 
verifying the presence of baryons in voids.
Penton \etal\ (2004) found that $29 \pm 4$\% of cosmological baryons 
reside in the low-$z$ \lya\ forest, while statistical studies 
(Penton, Stocke, \& Shull 2002) of 81 GHRS-discovered \lya\ absorbers 
showed that $22 \pm 8$\% of these \lya\ absorbers reside within voids, 
defined as regions located $>3h^{-1}_{70}$ Mpc away from the nearest 
known bright galaxy, in surveys complete to at least $L^*$ galaxy
luminosities.  

Our most recent survey (Danforth \& Shull 2007) examined the combined FUSE 
and STIS echelle archives, identifying 651 \lya\ absorbers along 28 
sight lines.  These absorbers and their metal lines can now be compared 
to existing galaxy redshift surveys, allowing us to identify void 
absorbers at $z \leq 0.1$. Our earlier work showed that absorbers in voids 
are preferentially found at lower column densities than absorbers in 
filaments, with few occurring at \NHI\ $\geq 10^{14}$ cm$^{-2}$.  
Integrating the absorber numbers (Fig.\ 12 in Penton, Shull, \& Stocke 
2002) using their standard assumptions for cloud sizes 
($100 h_{70}^{-1}$~Mpc) and ionizing intensities at 1 ryd
($J_0 = 10^{-23}$ erg~cm$^{-2}$ s$^{-1}$ Hz$^{-1}$ sr$^{-1}$), 
we estimated that $4.5 \pm 1.5$\% of all cosmological 
baryons are present deep within voids (Penton, Stocke \& Shull 2002; 
Stocke, Shull, \& Penton 2006).  If we assume the standard cosmological 
(Spergel \etal\ 2007) mixture of baryons 
($\Omega_b = 0.0455 h_{70}^{-2}$) and total matter 
($\Omega_m = 0.261 h_{70}^{-2}$) for Hubble constant 
$H_0 = (70~{\rm km~s}^{-1}~{\rm Mpc}^{-1}) h_{70}$, the fractional matter 
density deep within voids is then 
$\Omega_{\rm void} \approx 0.015 \pm 0.005$, consistent with numerical 
simulations of voids in $\Lambda$CDM universes (Gottl\"ober \etal\ 2003).  
Thus, the presence of \lya\ absorbers within voids appears to be
consistent with their formation by gravitational processes.
 
There are several possibilities for the metal enrichment of \lya\ clouds 
in voids.  Because metal-enriched bubbles from galactic winds are unlikely 
to fill the entire volume of intergalactic space (Ferrara, Pettini, \& 
Shchekinov 2000; Madau, Ferrara, \& Rees 2001), one might expect some regions
to be primordial, with essentially no metals.  Alternatively, the IGM may have 
been seeded with heavy elements from an early era of star formation,  
such as Pop~III stars in dwarf Primordial (dPri) galaxies (Ricotti,
Gnedin, \& Shull 2002, 2007). In this case, the dPri galaxies have now 
faded or lost their gas, and thus would no longer be visible.  Some evidence 
for the latter scenario is suggested by various searches for metals at 
$z \geq 5$ (Songaila 2001; Pettini \etal\ 2003) which find evidence for 
ubiquitous \CIV\ in \lya\ absorbers with \NHI\ $\geq 10^{14}$ \cd, 
at metallicities $Z_{\rm C} \approx 10^{-3.5}~Z_{\odot}$.  
Co-addition surveys (Cowie \& Songaila 1998) 
 of spectra of bright QSOs at $z \approx 2.5-3.0$ 
detected several heavy elements (\ion{C}{4}, \ion{O}{6}, \ion{Si}{4}) 
in \lya\ absorbers (e.g., Ellison \etal\ 1999, 2000;
Schaye \etal\ 2003; Aguirre \etal\ 2004; Simcoe \etal\ 2004).  Some of
these surveys conclude that large portions, and perhaps the entire 
high-$z$ IGM, is metal-enriched to $10^{-2}$ to $10^{-3}$ $Z_{\odot}$ 
by redshifts $z = 2-3$. Schaye \etal\ (2003) find that \CIV/\HI\ abundances 
depend on gas overdensity, which is correlated with column density \NHI.  
In their survey at $z \sim 2.5$, Simcoe \etal\ (2004) do not see a trend 
of decreasing metallicity toward lower-\NHI. Both [C/H] and [O/H] show 
log-normal distributions around mean values $\sim 10^{-2.85}\,Z_{\odot}$, 
but they find no universal metallicity floor.  Some 60--70\% of the 
absorbers are enriched to levels $Z \geq 10^{-3.5} Z_{\odot}$, but the 
remaining $\sim30$\% of lines have even lower (or undetectable) metal 
abundances.  Similar conclusions were found by Ellison \etal\ (1999). 

Ultraviolet spectra (HST, FUSE) allow us to test the two hypotheses 
for IGM metal injection by investigating the ``fossil record'' of the 
metallicity of \lya\ absorbers in low-$z$ cosmic voids.  This test is
possible because sites of recent star formation are not seen in their 
vicinity.  For example, McLin \etal\ (2002) found no galaxies within 
several hundred kpc of absorbers in very nearby voids ($cz\leq$3000 \kms) 
down to luminosity limits comparable to the luminosities of Local Group 
dwarf spheroidals.  Searches for \ion{H}{1} (21-cm) emission near \lya\
void absorbers have found no galaxies or galaxy-sized \ion{H}{1} clouds
(van Gorkom, private communication) to sensitivities of
$M_{\rm HI} \leq 10^8~M_{\sun}$.
These are important constraints, since galactic outflows 
driven by 10--40 Myr bursts of star formation from high-$z$, low-mass
galaxies create fairly small bubbles.  Even after 100 Myr, the bubble 
sizes are less than 10--30 kpc for 100--300 \kms\ winds.  
High-speed winds (1000 \kms) from massive galaxies would affect 
considerably less than 1 Mpc after 1 Gyr. 

In \S~2, we define our sample of void absorbers, in \S~3 we present 
details of the spectral analysis, and in \S~4 we discuss the 
inferred metallicity limits.  In \S~5 we
present our conclusions and discuss the promise for improving these 
results after the planned September 2008 installation of the
{\it Cosmic Origins Spectrograph} (COS) on HST. \\ 

\section{THE SAMPLE OF VOID ABSORBERS}

We began our study with the 651 \lya\ absorbers at $z \leq 0.4$ 
and their associated metal lines (\OVI, \CIII, \CIV, \SiIII, \SiIV, \NV, 
\FeIII) from the survey (Danforth \& Shull 2007) of 28 AGN sight lines
observed by both FUSE and STIS echelle (E140M).  These data include 
identifications of the associated \OVI\ (1031.926 \AA), 
\CIV\ (1548.203 \AA), and \SiIII\ (1206.500 \AA) absorption. 
Upper limits can also be obtained for the void absorbers.  These 
three metal lines were judged to provide the best combination of 
elemental abundance and line strength to serve as probes of absorber 
metallicity. The far-UV absorption line \CIII\ (977.021~\AA) 
was not used, as it falls in the SiC (low-S/N) portion of the FUSE band.
We also did not explore \SiIV\ (1393.755~\AA) in the longer wavelength
portion of STIS/E140M.  Surveys of Si~IV $\lambda1393.755$ might be 
useful to accompany C~IV $\lambda 1548.2$ as a tracer of the Si/C 
abundance in absorbers with log\,$U \approx -1.6\pm0.4$.  By examining the
(Si/C) abundances and the Si~IV/C~IV line strengths, we estimate that 
\SiIV/\CIV\ absorption lines should have the ratio 
$[N \, f \lambda]_{\rm SiIV}/[N \, f \lambda]_{\rm CIV} 
\approx 0.32 f_{\rm SiIV}/f_{\rm CIV}$, where $f_{\rm SiIV}/f_{\rm CIV}$
is the ratio of ionization fractions.  The (Si/C) abundance ratio
in the IGM might be a factor two higher than the assumed solar ratio, 
Si/C$_{\odot} \approx 0.132$. However, for most values of the ionization 
parameter, \CIV\ should be a more sensitive probe of metals than \SiIV.  
 
To characterize the large-scale distribution of galaxies, we 
employed our merged galaxy catalog of $>1.3$ million galaxy redshifts 
(Stocke \etal\ 2006) from CfA, SDSS/DR6, 2dF, and 6dF surveys, 
used in our statistical study of \OVI\ absorber environments.  In the
current paper, we identified voids as regions along the sight line where 
the three-dimensional ``nearest galaxy distance" (NGD) is $>1.4h^{-1}_{70}$ 
Mpc when the galaxy survey was complete to approximately $L^*$.  
We relaxed this criterion from our original NGD minimum distance,
from 3.0 to $1.4 h^{-1}_{70}$~Mpc, in order to search for more potential
void absorbers. Given the magnitude limits of many current galaxy 
surveys, this effectively limits our sample to $z_{\rm abs}<0.1$.  To
compute the line-of-sight absorber/galaxy distances, we assumed a
``retarded Hubble-flow model'' (Penton \etal\ 2002; McLin \etal\ 2002)
in which velocity differences $<300$ \kms\ are ignored and such objects 
are placed at the same distance from Earth. 

Within our void-absorber sample, we obtained no individual or co-added 
detections at any value of NGD. Note that our definition of voids is considerably 
more restrictive than that employed by other investigators 
(Rojas \etal\ 2005).  However, the number of usable void absorbers remains 
small (83 at $z < 0.1$) and varies for different ions, owing to the
limited spectral coverage, varying data quality, and occasional confusion 
with absorption lines of other species at other redshifts 
(Galactic \lya\ and metal lines).
We gave further consideration to those absorbers that provide column density
limits well below the metal-line detections in the large surveys
(Danforth \etal\ 2006; Danforth \& Shull 2007).  This requirement reduced the
number of absorbers to 61, only 12 of which have high-quality data for all three
ions of interest.  The strongest four H~I void absorbers (Table 1) dominate
our analysis.  Except for the 5040 \kms\ absorber in the PG\,1211+143 sight
line (Tumlinson \etal\ 2005), the strongest low-$z$ void absorbers all lie
$>5h^{-1}_{70}$ Mpc from the nearest galaxy and are unquestionably deep in
voids.  Our metallicity limits would not change substantially if we removed 
from consideration the PG\,1211+143 absorber, or the other absorbers with
$1.4 < {\rm NGD} < 3 h_{70}^{-1}$~Mpc. 

In Table~1 we list by column: (1) the void absorber sight line; 
(2) heliocentric absorber redshift; (3) NGD in $h^{-1}_{70}$~Mpc; 
(4) completeness limit of the galaxy survey work around this absorber
in $L^*$ units;
(5) column density log\,\NHI\ of the void absorber;
and (6--8) upper limits on metal-ion column densities.  
Column (9) lists S/N per resolution element at the location of 
each ion. Dashes in columns 6, 7, and 9 indicate that high-quality
data is not available for those ions in that specific absorber.
For example, the $z = 0.0168$ absorber toward PG~1211+143 has \OVI\
obscured by Galactic H$_2$ (Tumlinson \etal\ 2005).     
A complete table with similar data on all 61 void absorbers in our 
sample is available upon request. \\ 

\section{SPECTRAL ANALYSIS} 

In our analysis of the \CIV, \OVI, and \SiIII\ spectra, we considered only 
those absorbers for which the data are sufficiently good to yield 3$\sigma$ 
non-detections with log\,\NCIV\ $<12.8$, log\,\NOVI\ $<13.2$,
and log\,\NSiIII\ $<11.8$ in one or both metal lines.  These thresholds
were determined by examining the distribution of detected column densities  
and clean (S/N-limited) upper limits in Danforth \& Shull (2007)
from which these absorbers were drawn.  We made additional cuts from the 
sample to account for ambiguous void status (e.g., Ton\,S180 at 
$cz \sim 13,000$ \kms; Stocke \etal\ 2006\ and PKS\,0405-123 at 
$cz \sim 29,000$ \kms; Prochaska \etal\ 2006) and for potential confusion 
with absorption lines 
unrelated to the void absorbers.  After this culling, the total number of 
\lya\ absorbers was: \ion{O}{6} (20), \ion{C}{4} (49), and
\ion{Si}{3} (40).  Twelve absorbers, including the 3C\,273 entry in Table~1, 
have clean non-detections in all three ions.

The STIS/E140M and {\it FUSE} data were normalized in 10~\AA\ segments,
using low-order Legendre polynomial fits to the line-free continuum 
regions; see additional discussion in Danforth \etal\ (2006) and 
Danforth \& Shull (2007).  Absorption features
from unrelated IGM systems, ISM lines, and instrumental artifacts were
identified interactively and masked for each segment.  The signal-to-noise 
ratio (per resolution element) was found to be close to 
$S/N=1/\sigma(f_{\rm cont})$, where $\sigma(f_{\rm cont})$ is the 
standard deviation in the normalized continuum. 
 
It might seem advantageous to combine data from both lines of 
the \ion{O}{6} and \ion{C}{4} doublet.  However, the benefit from 
this procedure is, at best, marginal for unsaturated lines.  For 
Poisson noise, in a doublet with 2:1 ratio of oscillator strengths,  
this procedure increases the S/N by a factor $1.5/\sqrt 2$, or only 6\%.  
In practice, combining the data often degrades the S/N below that of 
the stronger doublet line alone; sometimes the result is degraded by 
20\% or more.  We have therefore chosen not to attempt these co-additions.  
Instead, each ion absorber is represented by a single optical depth vector,
$\tau_{\rm ion}(v) = - {\rm ln}\,[I(v)/I_c]$, which we combine using a 
pixel-by-pixel, S/N-weighted average of the unmasked pixels.  Because 
the original conversion from flux to optical depth is logarithmic 
rather than linear, the mean of the combined $\tau$ vectors is 
slightly greater than zero, $\langle\tau\rangle\sim0.01$, which we 
renormalize such that $\langle\tau\rangle=0$.  We converted the combined 
optical-depth vector to a normalized flux, $f(v)/f_c(v) = \exp [-\tau(v)]$, 
and calculated the combined (S/N)$_{\rm coadd} = 1/ \sigma(f_{\rm coadd})$.  
Because we see no evidence for significant non-Gaussian errors in 
either STIS or FUSE data, we characterize the S/N by a Gaussian $\sigma$.

Since no absorption features were apparent in the combined data, we
determined the $3\sigma$ limits on equivalent width from the total
S/N: $W_{\rm max} = (3\,\lambda_0 /(R~[S/N])$, where the spectral
resolution, $R =\lambda/\Delta\lambda$, is taken to be 
$R_{\rm STIS} \approx 42,000$ for STIS/E140M (\ion{C}{4} and \ion{Si}{3}) 
and $R_{\rm FUSE} \approx 15,000$ for FUSE (\ion{O}{6}).  Since the optical 
depth upper limit lies on the linear region of the curve of growth, the 
column density limit can be calculated from the linear formula,
$[N_{\rm ion}/10^{12}~{\rm cm}^{-2}] = [W_{\lambda}/8.85~{\rm m\AA}]
[f~\lambda_{1000}^2]^{-1}$, where $f$ is the absorption oscillator
strength and $\lambda_{1000}$ is the wavelength in units of 1000~\AA.   
For the three metals ions of interest, the parameters 
({\it Ion}: $f$, $\lambda_{1000}$) have the values:  
(\OVI: 0.133, 1.032), (\CIV: 0.190, 1.548), (\SiIII: 1.70, 1.206).   
Thus, in units of $10^{12}N_{12}$ \cd, we have \wlam(\CIV) = (4.03~m\AA)$N_{12}$,   
\wlam(\OVI) = (1.25~m\AA)$N_{12}$, and \wlam(\SiIII) = (21.9~m\AA)$N_{12}$.     
We can now compare our void metal-line upper limits with the available
\NHI\ data.  Since the spectra were combined using a S/N-weighted
average, we perform a weighted average of \NHI\ to determine a 
total \NHI, but weighted by metal-line S/N for each ion.  Thus, the final 
value of \NHI\ differs for each sample and each ion.  
In Table 1, we list column densities \NHI\ and upper limits for 
$N_{\rm ion}$ in the four strongest void absorbers.

For 49 void \ion{C}{4} absorbers, we obtain a combined S/N = 84.5 per 
resolution element, which translates to $W_{3\sigma}(1548) <1.3$~m\AA\ and 
log\,\NCIV\ $<11.51$. Similarly, for 20 \ion{O}{6} void absorbers, we obtain
a combined S/N = 48.5 yielding $W_{3\sigma}(1032) < 4.3$~m\AA\ and 
log\, \NOVI\ $<12.53$.  Finally, the 40 \SiIII\ void absorbers combine 
to give S/N = 84.8 so that $W_{3\sigma}(1206) < 1.0$~m\AA\ and 
log\,\NSiIII\ $< 10.67$. 
The S/N-weighted mean \HI\ column densities for these three samples are:
log\,\NHI\ = 13.46, 13.47, and 13.55 for the \CIV, \OVI, and \SiIII\ 
samples, respectively.  These translate into mean upper limits for ion ratios 
of: \NCIV/\NHI\ $<0.0112$, \NOVI/\NHI\ $<0.115$, and 
\NSiIII/\NHI\ $<0.0013$.  For reference, in our low-$z$ survey of O~VI 
(Danforth \& Shull 2005), the detected \OVI\ absorbers typically had ion
ratios ranging from \NOVI/\NHI\ $\approx$ 0.02 to 2.  This range
was interpreted to indicate a wide spread in multiphase ionization 
conditions of hot and photoionized gas in galaxy filaments. 

We also measured integrated equivalent widths across the possible locations 
of the coadded void metal-line absorption. Most detected metal-line 
features have FWHM $<50$ \kms, so we integrate the coadded $\tau$
and equivalent width $W$ vectors over a range, $v_{\rm HI} \pm 50$ \kms,
of ion wavelengths expected from $v_{\rm HI}$, the \lya\ radial 
velocity.  The standard deviations in the equivalent width measurements 
correspond to $3 \sigma(W_{\rm int})$ of 0.4~m\AA, 1.5~m\AA, and 
0.3~m\AA, for \CIV, \OVI, and \SiIII, respectively, consistent with 
continuum regions measured in 100 \kms\ boxes elsewhere in the coadded data.  
They are also consistent with the S/N-derived equivalent width limits above.

Owing to the improved S/N of the coadded void absorbers, our coadded
ion ratios, $N_{\rm ion}/N_{\rm HI}$, are more stringent than those 
obtained from the individual absorbers in the survey.  Table~1 shows 
that some individual absorbers have even lower ion ratios, owing to 
a few unusually strong \ion{H}{1} lines (\NHI\ $\geq 10^{14}$ \cd).  
Although the full sample co-addition gives the best indication of the 
ion ratios for the full ensemble of void
absorbers, the most stringent lower limits come from using only the
few high column density absorbers in Table~1.  This is explicitly
shown in Table~2, which lists ion ratios for several sub-samples of void 
absorbers.  The {\it ``Best Individual''} absorber is the one whose 
limits yield the smallest ratio.  The {``Top Few''} sample derives ion ratios 
from the coaddition of the strongest \ion{H}{1} systems (1, 3, or 4 for
\OVI, \CIV, and \SiIII, as listed in Table~1).  In this method, we
perform a pixel co-addition for the metal lines and adopt a S/N-weighted
\NHI, assuming that all void absorbers all drawn from the same homogeneous
sample with a mean and variance in metallicity.   
The {\it ``Complete Sample''} refers to those 12 absorbers with
good data in all three ions, and the {\it ``All Available''} sample uses a
pixel coaddition of all 61 void absorbers, although the total used for
each ion is somewhat less (20 \ion{O}{6}, 49 \ion{C}{4} and 40 \ion{Si}{3}).  

Table~2 shows that there are no strong trends in ion ratios for these
different sub-samples.  We do see small sample-to-sample differences,
largely reflecting the mean \NHI\ and spectral S/N of these samples. 
Therefore, we have chosen to apply the limits on metals obtained 
from the {\it ``Top Few''} sample as indicative for the complete sample, since 
there is no evidence that these differences result from anything more than 
the S/N available in these spectra. Therefore, from Table~2 we choose for  
further analysis: \NCIV/\NHI\ $<0.0063$, \NOVI/\NHI\ $<0.045$, and  
\NSiIII/\NHI\ $<0.00085$.  The co-addition method employed here assumes 
that void absorbers are selected from a uniform population in metallicity.  
Thus, the upper limits on
\CIV\ and \SiIII\ from the {\it Top Few} sample, derived from pixel
co-addition, are 22--26\% lower than from the best individual absorber 
data. \\ 

\section{IONIZATION ANALYSIS}

In order to determine limits on elemental abundances from our limits on 
individual ion abundances, we must apply ionization corrections.  
For example, the conversion from a limit on [OVI/HI] to a limit on
[O/H] requires knowledge of the physical state of the gas along the 
line of sight to the background quasars.  Most measurements and models 
(Danforth \& Shull 2005, 2007; Simcoe \etal\ 2004) suggest that the 
IGM absorbers consist of ``multiphase gas", in which \OVI\ and other
high ions arise in hot, collisionally ionized gas ($10^{5-6}$~K), 
while \HI\ and significant portions of \CIII, \SiIII, \CIV, and \SiIV\ 
exist in warm ($10^{4.0-4.5}$~K) photoionized gas.  Many of these
absorbers appear to be kinematically associated within $\pm20$~\kms, 
but this many not be surprising given the large spatial extents
(100-200 kpc) inferred for \lya\ absorbers.   Because we do not possess
detailed knowledge about the cloud geometries or physical structures, 
we introduce simplified models for the ionization state of the gas, 
which provide a representative range of metallicity limits.

We begin our analysis with the case of photoionization, which is most 
appropriate for \CIV\ and \SiIII.  Collisional ionization will be discussed
later. We have constructed several grids of ionization equilibrium models
using the photoionization code Cloudy (version 07.02, Ferland \etal\ 1998).
We assume that the gas is optically thin and photoionized by a quasar-dominated
metagalactic ionizing background (Haardt \& Madau 1996, 2001), with 
specific intensity, $J_{\nu} = J_0 (\nu/\nu_0)^{-\alpha}$.
The effective spectral index, $\alpha$, in the Lyman continuum is produced 
by the range of intrinsic source spectra, altered by transmission and
reprocessing through the intervening IGM (Fardal \etal\ 1998; Shull \etal\ 1999).    
Observations of low-redshift AGN (Telfer \etal\ 2002) suggest intrinsic
spectral indices ranging from $0 < \alpha_s < 3$, with a mean 
$\langle \alpha_s \rangle$ lying somewhere between 1.5 and 2.0.   
Because transmission tends to harden the radiation spectrum ($\alpha < \alpha_s$),
we scale our results to an effective $\alpha \approx 1.5$ and normalize  
the ionizing background at 1 ryd to 
$J_0 = 10^{-23}\,J_{-23}~{\rm erg~cm}^{-2}~{\rm s}^{-1}\,{\rm Hz}^{-1}\,{\rm sr}^{-1}$, 
close to the derived value at $z \approx 0$ (Shull \etal\ 1999).  When the 
radiation intensity is fixed, variations in the ionization parameter, 
$U \propto J_0 / n_H$, represent primarily variations in the hydrogen 
density $n_H$. 

To check these effects and their analytic scaling,
we examine the following formulae. For a diffuse, isotropic radiation 
field, $J_{\nu}$, incident on optically thin \lya\ clouds from all directions, 
we can write down formulae for the density, $n_{\gamma}$, of ionizing photons, 
the density, $n_H$, of hydrogen nuclei, the hydrogen photoionization rate, 
$\Gamma_H$, and the photoionization parameter, $U = n_{\gamma}/n_H$:  
\begin{eqnarray}
 n_{\gamma} &=& \frac {4\,\pi \, J_0}{hc\,\alpha}=(4.22\times10^{-7}~{\rm cm}^{-3}) 
                J_{-23} \left( \frac {1.5} {\alpha} \right) \\
 n_H       &=& \frac {\rho_{\rm cr} (1-Y_p) \Omega_b} {m_H} =
             (1.89 \times 10^{-7}~{\rm cm}^{-3}) \, \delta_H   \\
  U         &=& \frac {n_{\gamma}} {n_H} = (2.23) J_{-23} \left( 
                 \frac {1.5} {\alpha} \right) \, \delta_H^{-1} \\ 
 \Gamma_H &\approx& \left[ \frac {4\,\pi \, J_0 \, \sigma_0} {h\,(\alpha+3)} \right] 
                = (2.66 \times 10^{-14}~{\rm s}^{-1}) J_{-23} \left[ 
                     \frac {4.5} {\alpha + 3} \right] \; .
\end{eqnarray} 
Here, we have written the ($z=0$) IGM hydrogen density, $n_H$, in terms of the 
baryon mass density ($\rho_{\rm cr} \, \Omega_b$), corrected for 
primordial helium abundance, $Y_p \approx 0.2474$, and scaled to the
hydrogen overdensity parameter $\delta_H$.  The \HI\ photoionization  
rate involves frequency-integration of $J_{\nu}$ times an approximate 
photoionization cross section, $\sigma_{\nu} \approx \sigma_0 (\nu/\nu_0)^{-3}$, 
where $\sigma_0 = 6.30 \times 10^{-18}$ cm$^2$ at threshold $h\nu_0 = $1 ryd.
From their IGM simulations, Dav\'e \etal\ (1999) found an empirical relation,
$\delta_H \approx 20 N_{14}^{0.7} \, 10^{-0.4z}$, between overdensity 
$\delta_H$ and column density, $N_{\rm HI} = (10^{14}~{\rm cm}^{-2}) N_{14}$.
A similar relation was found by Schaye (2001).
We assume photoionization equilibrium, with case-A recombination rate
coefficient, 
$\alpha_H^{(A)} = (2.5 \times 10^{-13}~{\rm cm}^3~{\rm s}^{-1}) T_{4.3}^{-0.76}$, 
scaled to temperature $T = (10^{4.3}~{\rm K})T_{4.3}$. We can then derive the 
photoionization corrections for the \HI\ neutral fraction, $f_{\rm HI}$, 
and estimate the characteristic absorber sizes, $D_{\rm cl}$: 
\begin{eqnarray}
  f_{\rm HI} &=& \frac {n_{\rm HI}}{n_H} = 
     1.164 \left[ \frac {n_H \alpha_H^{(A)}(T)} {\Gamma_H} \right]   
     \approx (11.7) n_H \, T_{4.3}^{-0.76} \, J_{-23}^{-1} \; , \\ 
  D_{\rm cl} &\approx& \frac {N_{\rm HI}} {n_H \, f_{\rm HI} } 
     \approx (0.194~{\rm Mpc}) \, N_{14}^{-0.4} \, T_{4.3}^{0.76} J_{-23} \; .
\end{eqnarray}  
Thus, the observed four strong \lya\ absorbers, with log\,\NHI\ $= 14.0-14.63$,
are predicted to have overdensity factors $\delta_H = 20-55$ and characteristic
sizes $D_{\rm cl} =$ 100-200 kpc.  These ranges are probably uncertain to a factor
of two. We can conveniently express the photoionization parameter in terms of 
the parameters, $J_{-23}$ and $\delta_H$:
\begin{equation}
  U = (0.045) \, J_{-23} \, \left( \frac {1.5} {\alpha} \right) 
     \left( \frac {50} {\delta_H} \right)  \; .
\end{equation}

\begin{figure}[t]
  \epsscale{2.00}
  \plottwo{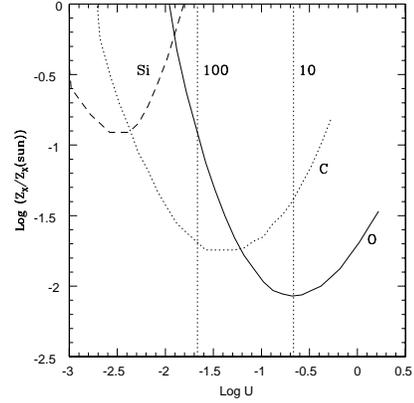}{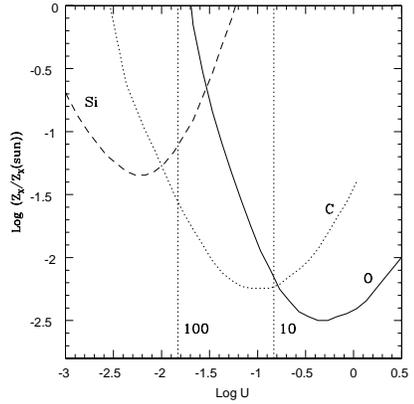} 
\caption{Limits on metallicity ($Z/Z_{\odot}$) vs.\ photoionization
  parameter ($U$) for silicon, carbon, and oxygen, using observed limits on
  \SiIII, \CIV, and \OVI\ from Table 1.  Ionization corrections were derived 
  from photoionization models of Ferland \etal\ (1998).  
  log\,$U = -1.6 \pm 0.4$ for low-$z$ absorbers (Danforth \& Shull 2007).  
  (Left) Assumes hard ionizing AGN continuum with source spectral index 
  $\alpha_s = 1.8$ plus IGM transmission.
  (Right) Assumes mixture of AGN and softer (starburst) continuum.  
  The two vertical dashed lines show overdensities ($\delta_H$ = 10 and 100) 
  corresponding to the strong \lya\ absorbers studied here, with fixed 
  $J_{-23} = 1$.  Metallicity limits, based upon non-detections of \CIV\ and 
  \OVI, are $Z < 10^{-1.8 \pm 0.4} Z_{\odot}$, assuming log\,$U = -1.6 \pm 0.4$ 
  for low-$z$ absorbers (Danforth \& Shull 2007) and including systematic 
  uncertainties of the ionizing spectrum.}  
\end{figure}

We assume that the absorbers have uniform gas density, $n_H$, that the 
diffuse ionizing radiation, $J_{\nu}$, impinges on the absorbers uniformly
from all directions, and that the relative metal abundances are the same
as solar photospheric ratios given by Asplund \etal\ (2005), with
(C/H)$_{\odot} = 10^{-3.61}$, (O/H)$_{\odot} = 10^{-3.34}$, and 
(Si/H)$_{\odot} = 10^{-4.49}$.  Previous groups used older abundances 
(Anders \& Grevesse 1989; Grevesse \& Sauval 1998) in which these
abundances were somewhat higher (C/H = 331 ppm and O/H = 676 ppm).   
If we had used these older values, our limits on metal abundances would 
be 0.17 dex lower for oxygen and 0.13 dex lower for carbon, while the 
silicon limit would remain unchanged. For a given value of $U$, we vary 
the global metallicity until we achieve the highest value compatible with 
the non-detections of \SiIII, \CIV, and \OVI\ associated with the appropriate \NHI.  

Figure 1 shows the constraints on [Si,C,O] metallicity for a given photoionization 
parameter associated with upper limits on [\SiIII, \CIV, \OVI].  As $U$ increases, 
the density $n_H$ decreases for fixed $J_0$, and the upper limit provided by the 
higher ion stages provides the stronger constraint. The vertical lines in the
plot correspond to hydrogen densities, $n_H$, assuming an ionizing background 
$J_{-23} = 1$.  Our best upper limits on metal absorption 
lines come from the four specific absorbers, with \NHI\ ranging from 
$10^{14.0}$ to $10^{14.63}$ cm$^{-2}$, corresponding to  
$\delta_H \approx$ 20--55 in the current epoch.   For a mean IGM density
$\langle n_H \rangle = 1.89 \times 10^{-7}$ cm$^{-3}$ at $z \approx 0$, the 
vertical lines in Figure 1 correspond to a somewhat wider range of overdensities 
$\delta_H =$ 10--100. The best limits on void metallicities come from the 
absence of \CIV\ and \OVI, where we adopt a range of photoionization parameter, 
log\,$U \approx -1.6 \pm 0.4$, inferred from the low-$z$ IGM ionization ratios 
of \SiIII/\SiIV\ and \CIII/\CIV\ found by Danforth \& Shull (2007). 
The two panels (Fig.\ 1a,b) show the approximate range of uncertainty in
metallicity introduced by the likely range in both $J_{\nu}$ and log\,$U$. 

The absorbers with lower \NHI, listed in Table 2 as the {\it ``All Available''} 
sample with log $\langle N_{\rm HI} \rangle = 13.5$ (and thus 
log\,$U \approx -1$) have metallicity limits set by the absence of O~VI, even for 
photoionized gas. However, if $J_0 < 10^{-23}$ in voids, or if the radiation 
field is softer than a Haardt-Madau spectrum, the best void metallicity limits 
are set by \CIV.  Unfortunately, at low redshift, there are no strong 
constraints on $J_\nu$ in voids.  From Figure 1 we find 
$Z < 10^{-1.8 \pm 0.4} Z_{\odot}$, for log\,$U = -1.6 \pm 0.4$.
The errors include systematic uncertainties in the assumed ionizing 
radiation field. 

A similar analysis, using \OVI\ in collisionally-ionized gas, yields 
metallicity limits 0.5--1.5 dex lower, depending upon the assumed post-shock 
temperature of the gas.  More likely, \OVI\ is predominantly collisionally
ionized, and the \HI\ resides in photoionized gas.  Measurements and models
(Danforth \& Shull 2005, 2007; Simcoe \etal\ 2004) suggest that the
IGM absorbers consist of ``multiphase gas", in which \OVI\ and other
high ions arise in hot, collisionally ionized gas ($10^{5-6}$~K),
while \HI\ and significant portions of \CIII, \SiIII, \CIV, and \SiIV\
exist in warm ($10^{4.0-4.5}$~K) photoionized gas.  Many of these
absorbers appear to be kinematically associated within $\pm20$~\kms,
but this many not be surprising given the large spatial extents
(100-200 kpc) inferred for \lya\ absorbers.  
Numerical simulations of low-$z$ \lya\ absorbers find 
that absorbers far from galaxies are dominated by photoionization (Dav\'e \etal\
1999), with collisionally-ionized absorbers found much closer to galaxies.  
In these models, over 80\% of the absorbers more than 1 Mpc from 
any galaxy are photo-ionized. Thus, the case of photoionization places the 
least restrictive and most appropriate limits on metals in voids. \\ 
 
\section{CONCLUSIONS AND FUTURE PROSPECTS}

We have used a large database (Danforth \& Shull 2007) of 651 \lya\ 
absorbers found in both STIS/E140M and FUSE spectroscopy to investigate 
the metallicity of gas in cosmic voids.  We cross-correlated a merged 
galaxy catalogue with $>$ 1.2 million entries with the \lya\ absorber 
locations to identify a sample of 61 ``void absorbers" at $z < 0.1$, 
located $>1.4 h^{-1}_{70}$ kpc from the nearest L$^*$ or brighter galaxy. 
Using both individual spectra of high column density 
(\NHI\ $>10^{14}$ cm$^{-2}$) absorbers and pixel-additions, 
we find equivalent width limits (\OVI, \CIV, \SiIII) of 1--4 m\AA. 
The best metallicity limits are set from the strongest 4 absorbers 
in voids, all of which have \NHI\ $>10^{14}$ cm$^{-2}$ (see Tables 1 and 2). 
With ionization corrections from standard photoionization models, 
the metallicity limits (3$\sigma$) in voids are 
$Z < 10^{-1.8 \pm 0.2} Z_{\odot}$ for a range of ionization parameters 
(log\,$U = -1.6 \pm 0.4$) that characterize the low-\NHI\ IGM at $z < 0.1$ 
(Danforth \& Shull 2007). 

Collisionally ionized gas models set even lower metallicity limits, based 
on the non-detection of \OVI.  While the results from this first study 
indicate that cosmic voids contain gas that is quite metal poor, the limit 
of $Z < 10^{-1.8} Z_{\odot}$ by itself is insufficient to constrain the 
existence of an early burst of star formation within voids. Extremely 
metal poor (EMP) stars in our own Galaxy have [Fe/H] $\leq -3.0$ (Cayrel \etal\ 
2004) in Fe-peak elements, which in some sense sets an upper limit on the 
metallicity of the earliest stars.  However, many of these EMP stars have
enhanced [C/Fe] and [O/Fe] (Christlieb \etal\ 2002), so the actual metallicities  
may exceed 1\% solar.  Simulations by Ricotti, Gnedin, \& Shull (2002)
suggest that IGM metal abundances in the IGM can rise to 
$\sim 10^{-3.5} Z_{\odot}$ by $z\sim$10.  Assuming standard metal yields,
a co-moving star-formation-rate (SFR) density of 
$\rho_{\rm SFR} \approx 0.1M_{\odot}~{\rm yr}^{-1}~{\rm Mpc}^{-3}$,
typical of the peak values at $z \approx 6$, will produce 1\% solar 
metallicity in $\sim$1 Gyr (Tumlinson \etal\ 2004). 
The simulations also find an exponentially increasing co-moving rate of 
star formation from $z = 20$ to $z = 10$. The early burst of metal
production is likely to occur shortly after $z \approx 10$, when 
simulations suggest that $\rho_{\rm SFR}$ peaks at 
$0.1M_{\odot}~{\rm yr}^{-1}~{\rm Mpc}^{-3}$.  One scenario is that a 
few generations of stars formed from low-mass (dPri) galaxies within voids, 
and then star formation ceased from $z \sim 10$ until the current epoch.
In this case, the level of metallicity, $Z \approx 10^{-5} Z_{\odot}$ 
(Ricotti, Gnedin, \& Shull 2007), would be well below the detection 
threshold of the current best limits at both high and low redshift.  

On the other hand, the mean metallicity of \lya\ absorbers in galaxy 
filaments today has been estimated at $Z \approx 0.1 Z_{\odot}$ 
(Danforth \& Shull 2005; Stocke \etal\ 2006), nearly an order of magnitude 
larger than the $10^{-1.8} Z_{\odot}$ limit derive for voids at $z \leq 0.1$.  
At much higher redshift, many groups have made detections 
or set upper limits on metals in the Ly$\alpha$ forest.  Songaila (2001)
found the (\CIV) metallicity at $z = 5$ to be $Z \geq 10^{-3.5} Z_{\odot}$,
while at $z =$ 2.5--2.8, Simcoe \etal\ (2004) found a mean metallicity of 
oxygen and carbon of $\sim 10^{-2.8 \pm 0.8} Z_{\odot}$, with $\sim$30\% of 
the absorbers at \NHI\ $\geq 10^{13.5}$ cm$^{-2}$ showing no detectable 
metals at all, $Z < 10^{-3.5} Z_{\odot}$. Ellison \etal\ (1999, 2000)
made similar explorations of weak \CIV\ absorbers down to 
log\,\NCIV\ $\approx 11.7$ and found no evidence for flattening of the
$N_{\rm CIV}^{-1.44}$ power-law distribution. Pettini \etal\ (2003)
measured 16 \CIV\ systems with log\,\NCIV\ $= 12.50-13.98$ at $z > 5$,
and deduced a comoving mass density $\Omega_{\rm CIV} =
(4.3\pm2.5) \times 10^{-8}$ for systems with log\,\NCIV\ $>13.0$.
This metal abundance is consistent with that measured at $z<4$ and
with the finding (Songaila 2001) that the \CIV\ column-density distribution
and inferred metallicity show little evolution over this period.     
 
Because void absorbers account for $\sim$22\% of all low-column density 
\lya\ absorbers in the current epoch (Penton, Stocke \& Shull 2002), 
it is tempting to identify the 30\% metal-free absorbers (Simcoe \etal\
2004) at $z\sim$2.5 with absorbers in voids.  Simcoe \etal\ (2004) 
use the Grevesse \& Sauval (1998) solar reference abundances, so our best 
limit is $Z \leq 10^{-1.9} Z_{\odot}$ from C/H in Figure~1. 
If the local void absorbers were drawn from the metallicity distribution 
at $z\sim2.5$ quoted above, then we should have detected some 
metal lines in our current voids sample. From the current evidence, it is 
easier to identify the metal-bearing population at $z \sim 2.5$ with the 
$10^{-1} Z_{\odot}$ absorbers in galaxy filaments today, so we suggest that 
IGM metallicity in filaments has increased by a factor of 30--100 over 
that time.   High-redshift absorbers without metals 
($Z \leq 10^{-3.5} Z_{\odot}$) probably reside in voids and have remained 
extremely metal poor or pristine even to the current epoch.
If this association of metal-poor absorbers and voids is correct, then 
the absorbers with $Z \leq 10^{-3.5}~Z_{\odot}$ should have a systematically
lower two-point correlation function amplitude than the metal-rich 
absorbers at comparable redshift.
 
Our comparisons between high-$z$ and low-$z$ absorbers were made for 
systems with similar \NHI.  However, simulations tell us that individual 
\lya\ absorbers evolve in \NHI\ over time.  Dav\'e \etal\ (1999) and 
Schaye \etal\ (2001) suggest a factor of $\sim20$ evolution in absorber 
numbers, from $z \sim3$ to the present; Penton, Stocke \& Shull (2004) 
measure a factor of 3--10 over the same range.
Therefore, the $z\sim2.5$ absorbers at log\, \NHI\ = 13.5--15.5 should 
become roughly an order of magnitude lower in \NHI\ in the present epoch. 
There appears to be some disagreement over the presence of
metallicity-density correlation in fits to [C/H] or [O/H]
with log\,\NHI.  Schaye \etal\ (2003) find a density dependence in their
\CIV\ survey, but Simcoe \etal\ (2004) suggest that these effects may
be masked by systematic uncertainties in the ionization radiation field.
The interpretation of these trends is tied to uncertainties in the
sources and spectral shapes of the UVX background.
(Giroux \& Shull 1997; Shull \etal\ 1999).
With no apparent dependence of metallicity with \NHI\ (Simcoe \etal\ 2004),
our metallicity comparisons remain approximately correct.  
 
While these conclusions are supported by the upper limits derived here for 
current-epoch void absorbers, they would be strengthened considerably by 
upper limits $\sim1$ dex lower. With the {\it Cosmic Origins Spectrograph} 
(COS) awaiting installation into HST, it is worthwhile to consider how these 
limits could be improved in the COS era. It will be hard to improve on the 
oxygen limit, as it is set by the non-detections of \OVI\ by FUSE.  
At $z \geq 0.12$, the \OVI\ lines shift into the HST/COS band,
but current galaxy survey work is of insufficient depth to determine the 
locations of voids unambiguously at $z \geq 0.1$.  Since the current 
metallicity limit is based 
largely upon just one \OVI\ absorber, finding $\sim10$ new candidate void 
absorbers at \NHI\ $\geq 10^{14}$ \cd\ and $z \geq 0.12$ seems possible,
and it could reduce the oxygen abundance limit significantly. 
Conducting the necessary deep galaxy surveys around a few 
of these sight lines also seems tractable. 

For \CIV\ to set more stringent metallicity limits than \OVI\ across the 
full range of log\,$U$ would require $\sim30$ void absorbers with 
\NHI\ $\geq10^{14}$ \cd\ selected from a large survey of low-$z$ sight lines.
Let us assume an absorber frequency, $dN/dz \approx 25$ (Penton \etal\ 2004)
for \lya\ absorbers with \NHI\ $\geq 10^{14}$ \cd, with a redshift pathlength 
$\Delta z \approx 0.1$ per target and a 10\% probability of finding 
strong \HI\ absorbers within voids.  Therefore, finding 30 strong void absorbers 
would require observing $\sim100$ targets. We could reduce this number
somewhat by taking longer exposures (higher-S/N spectra) sensitive 
to weaker metal lines in absorption systems with log\,\NHI\ $\geq 13.6$.   
An HST/COS survey would rely on current and ongoing galaxy surveys 
to find the voids. Such target lists exceed the number of AGN now 
envisioned in the COS Guaranteed Time Observer (GTO) program, but 
they could be accommodated in a COS Key Project on the IGM.
It would marvelous to find just one void absorber
with \NHI\ $\approx 10^{15}$ \cd, for which it might be possible to detect 
metals down to metallicities of $10^{-2.5} Z_{\odot}$ in these strong
ultraviolet resonance lines.  Given the sensitivity of the \OVI\ lines,
a far-UV mission replacing FUSE would be especially important, in order 
to search for weak, low-redshift \OVI\ and \CIII\ absorbers.  
The low-$z$ fossil record of metallicity may provide the best evidence
for dwarf Primordial galaxies and the first stars (Ricotti et al. 2007). \\ 

\acknowledgements
We acknowledge the financial support from NASA/HST guest observer grants 
HST-GO-06593.01, HST-GO-09506.01 and from the COS GTO Science Team.  
Additional support for IGM studies at the University of Colorado came from 
FUSE (NASA grant NNG06GI91G) and NASA Astrophysical Theory grant NNX07AG77G.   
JTS thanks Michael Vogeley and Fiona Hoyle for their invitation to speak
on this subject at conferences in Aspen and Amsterdam.


\begin{deluxetable}{llccccccc}
\tabletypesize{\scriptsize}
\tablecolumns{9}
\tablewidth{0pt}
\tablecaption{Four Strong Void Absorbers}
\tablehead{\colhead{Sight Line}         &
        \colhead{$z_{\rm abs}$}         &
        \colhead{NGD\tablenotemark{a}}  &
        \colhead{Limiting \tablenotemark{b}}&
        \colhead{log\,\NHI}            &
        \colhead{log\,$N_{\rm OVI}$\tablenotemark{c}}  &
        \colhead{log\,$N_{\rm CIV}$\tablenotemark{c}}  &
        \colhead{log\,$N_{\rm SiIII}$\tablenotemark{c}}  &
        \colhead{(S/N) per resel}              \\
        \colhead{}                      &
        \colhead{}                      &
        \colhead{$h^{-1}_{70}$\,Mpc}    &
        \colhead{$L(L^*)$}              &
        \colhead{(cm$^{-2}$)}           &
        \colhead{(cm$^{-2}$)}           &
        \colhead{(cm$^{-2}$)}           &
        \colhead{(cm$^{-2}$)}          &
        \colhead{1032/1548/1206}               }
\startdata
 3C\,273       & 0.06656 &  6.7 & 1    & $14.17\pm0.11$ & $<12.82$ &
       $<12.27$ & $<11.20$  &  ~20~~/~20~/~35 \\
 3C\,351       & 0.04086 & 10.0 & 0.3 & $14.63\pm0.14$ & \nodata  &
       $<12.56$ & $<11.67$  & \nodata/~10~/~12 \\
 PG\,1211+143  & 0.01680 &  1.8 & 0.5  & $14.00\pm0.15$ & \nodata  &
       $<12.24$ & $<11.46$  & \nodata/~21~/~19 \\
 PKS\,0312-770 & 0.07368 & 11.3 & 0.5  & $14.48\pm0.85$ & \nodata & \nodata &
      $<11.71$  & \nodata/\nodata/ 11 \\
\enddata
\tablenotetext{a}{NGD = Nearest Galaxy Distance (see text in \S~2)}
\tablenotetext{b}{Nearby galaxy survey completeness limit, in units of $L^*$,
   at absorber redshift}
\tablenotetext{c}{Upper limits ($3 \sigma$) of metal ion column densities.}
\end{deluxetable}


\begin{deluxetable}{lccccc}
\tabletypesize{\footnotesize}
\tablecolumns{6}
\tablewidth{0pt}
\tablecaption{Metal Ion Limits in Void Absorbers}
\tablehead{\colhead{Sample}   &
        \colhead{Sample} &
        \colhead{S/N-weighted} &
        \colhead{$N_{\rm OVI}/N_{\rm HI}$}    &
        \colhead{$N_{\rm CIV}/N_{\rm HI}$}    &
        \colhead{$N_{\rm SiIII}/N_{\rm HI}$}   \\
        \colhead{}       &
        \colhead{Size}   &
        \colhead{log\,$\langle N_{HI}\rangle$} &
        \colhead{}       &
        \colhead{}       &
        \colhead{}     }
\startdata
Best Individual   &   1/1/1  & 14.17/14.63/14.17 & $<0.045$ & $<0.0085$    &
                                                          $<0.0011$ \\
Top Few (Table~1) &   1/3/4  & 14.17/14.27/14.29 & $<0.045$ & $<0.0063$ &
                                                          $<0.00085$ \\
Complete Sample   &    12    & 13.59/13.49/13.57 & $<0.043$ & $<0.019$     &
                                                          $<0.0025$ \\
All Available     & 20/49/40 & 13.46/13.47/13.55 & $<0.115$ & $<0.0112$    &
                                                          $<0.0013$ \\
\enddata
\end{deluxetable}


\end{document}